\documentclass[seceq]{ptptex}
\usepackage{wrapft}
\usepackage{graphicx}

\pubinfo{Vol.~***, No.~*, *** 2003}

\title{
Present Status of the Theory of Fission of hot Nuclei
}

\author{
Peter \textsc{Fr\"obrich}$^{1,2,}$\footnote{E-mail: froebrich@hmi.de}
}

\inst{
$^1$Hahn-Meitner Institut Berlin, Theoretical Physics (SF5),
Glienicker Stra{\ss}e 100, 14109 Berlin, Germany\\
$^2$Institut f\"ur Theoretische Physik, Freie Universit\"at Berlin, Arnimallee
14, 14195 Berlin, Germany
}

\abst{
Recent progress in the theory of fission of hot nuclei is reported. We discuss
in particular  the properties of the friction form factor as function of the
deformation (and possibly of the temperature) which are necessary to
reproduce data concerning fission of hot nuclei and its accompanying light
particle and $\gamma$-ray emission. Recent theoretical work gives support to a
phenomenological friction form factor (proposed some time ago \cite{FGM93}),
which is weak for compact shapes and increases on the way to scission. }

\begin{document}

\maketitle

\section*{1.Introduction}

Fission is one of the main decay channels after fusion of two heavy ions.
The following discussion is restricted mainly to systems for which Bohr's
hypothesis is valid, which states that the formation and decay of the compound
nucleus are independent processes, i. e. fission starts only after a completely
equilibrated compound nucleus has been formed. We do not deal with systems
having a considerable fast fission component or with systems where
the dynamics from capture to the formation of the compound nucleus plays a
role. These topics are elsewhere discussed in these conference proceedings, see
the contributions of Hinde, Trotta, Hannape, Abe, Aritomo and Rummel.

In the following, we report on progress made since the
the review article \cite{FG98} appeared in
our understanding of fission of hot
nuclei and its acompanying processes (light particle and $\gamma$-ray
emission).
Whereas there is no doubt that a Langevin description plus a
Monte Carlo treatment of the evaporation processes provides the most
adequate dynamical description (one obtains physical insight by e.g.
sampling distributions which tell  which
particle with a particular energy and position is emitted along the
fission path at a particular time),  there is less agreement on the input
quantities which
enter the description. (i) A fusion model has to be applied in order to obtain
the initial spin distribution for fission. (ii) A choice of the relevant
variables has to be made for the shape parametrization with which the driving
force is constructed (it is not yet common practice to use the
free energy or the entropy). (iii) An evaporation model has to be coupled to
the dynamics. (iv) After reaching a stationary value for the fission width, a
modified
statistical model has to be added for computational reasons. (v) Most
controversial is the choice
of the friction form factor, the discussion of which is the main topic of this
article. We start in Section 2 with reporting on multi-dimensional Langevin
calculations\cite{NAK02}, which solve a long-standing problem concerning the
width of the kinetic energy distribution of the fission fragments. In Section 3
we discuss a chaos-weighted wall formula which supports the
phenomenological friction form factor proposed in Ref.\cite{FGM93} (see also
\cite{FG98,FL96}) in order to
reproduce simultaneously data for fission probabilities and pre-scission
neutron
multiplicities, which was not possible within a statistical model. This
friction form factor was subsequently used extensively to analyse experimental
data on pre-scission neutron multiplicities and fission (respectively survival)
probabilities, pre-scission neutron energy spectra, pre-scission charged
particle
(p,$\alpha$) and giant-dipole $\gamma$ multiplicities and spectra, evaporation
residue cross sections, fission times, temperatures at scission, and the
fission angular distribution; see the review article \cite{FG98}.
In Section 4 we turn to a discussion of a possible temperature dependence of
the friction form factor, and make some remarks on quantal corrections in
Section 5 before concluding.

\section*{2. The kinetic energy distribution of fission fragments }

There have been attempts to explain the measured correlation between the
kinetic energy distribution of fission fragments and pre-scission neutron
multiplicities with two dimensional (elongation and constriction) Langevin
dynamics by Wada et al. \cite{WAC93}, who use one-body dissipation (wall
friction) and by Tillack et al. \cite{TRS92}, who use two-body viscosity.
%
\begin{figure}[htb]
\begin{center}
\protect
\includegraphics*[bb= 70 300 500 650,
angle=0,clip=true,width=8cm]{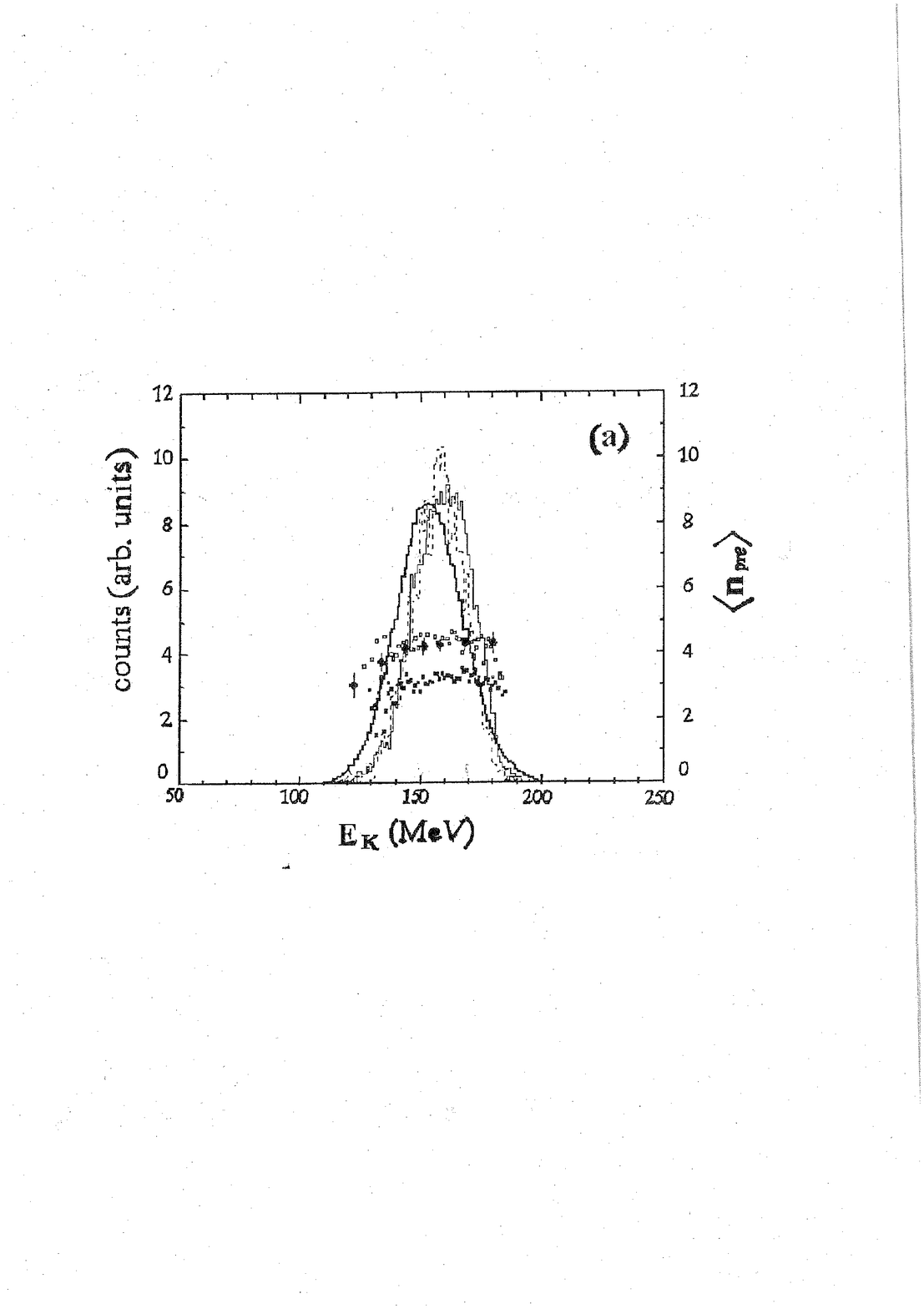}
\protect
\caption{The measured kinetic energy distribution (fat solid histogram) in
coincidence with the pre-scission neutron multiplicities (filled circles) are
compared with the corresponding calculated quantities with reduced widow
friction: $K_s=0.25$ (dashed histogram,open squares), and $K_s=0.5$ (thin solid
histogram, filled squares), from Ref.\cite{NAK02}.}
\end{center}
\end{figure}
In both investigations the width of the kinetic energy distribution came out by
about a factor of two too narrow. This was attributed to the neglect of the
mass asymmetry degree of freedom in some schematic calculations \cite{AARS96}.
Systematic multi-dimensional Langevin calculations \cite{NAK02} showed that
the recoupling of the mass asymmetry degree of freedom to the kinetic energy
distribution of the fission fragments leads to agreement with the measured
widths. For an example see figure 1.
There is agreement for `light' systems ($^{172}Y,^{205,215}Fr,^{224}Th$) with
the data for the TKE-widths and also for the pre-scission neutron
multiplicities. However, in order to find
agreement with experiment one has to reduce the strength of the wall
friction by a
factor $K_s\simeq 0.25-0.5$. For very heavy systems with $100\%$ fission
probability
such as $^{252,256}Fm$, the pre-scission neutron multiplicities cannot be
reproduced.
For systems with a long saddle to scission path, one seems to need a stronger
friction at large deformations. See the discussion in Section 3.

A reduction of the strength of the wall formula is also necessary for
reproducing experimental
fission times in a multi-dimensional Langevin description\cite{GMB02}.

The findings above are \underline{in contradiction} to the work of Abe and
collaborators \cite{WAC93,BAK03}, who apply the wall formula without reducing
its strength.

\section*{3. The chaos-weighted wall formula}

One of the essential assumptions in deriving the wall formula is that the
single-particle collisions with the wall are completely randomized. This
corresponds to chaotic motion. However the amount of chaos depends
on the deformation of the fissioning nucleus: in a spherical container one has
regular motion, with increasing deformation the fraction of chaotic motion
increases. This leads to a reduction of the wall formula due to incomplete
randomization. Pal and coworkers \cite{Pal96} have calculated a
factor taking into account the amount of chaos. This reduces the strength
of the wall formula as a function of the
deformation. This chaos-weighted wall friction form factor is shown in Fig.2 as
\begin{figure}[htb]
\begin{center}
\protect
\includegraphics*[bb= 40 90 480 780,
angle=-270,clip=true,width=11cm]{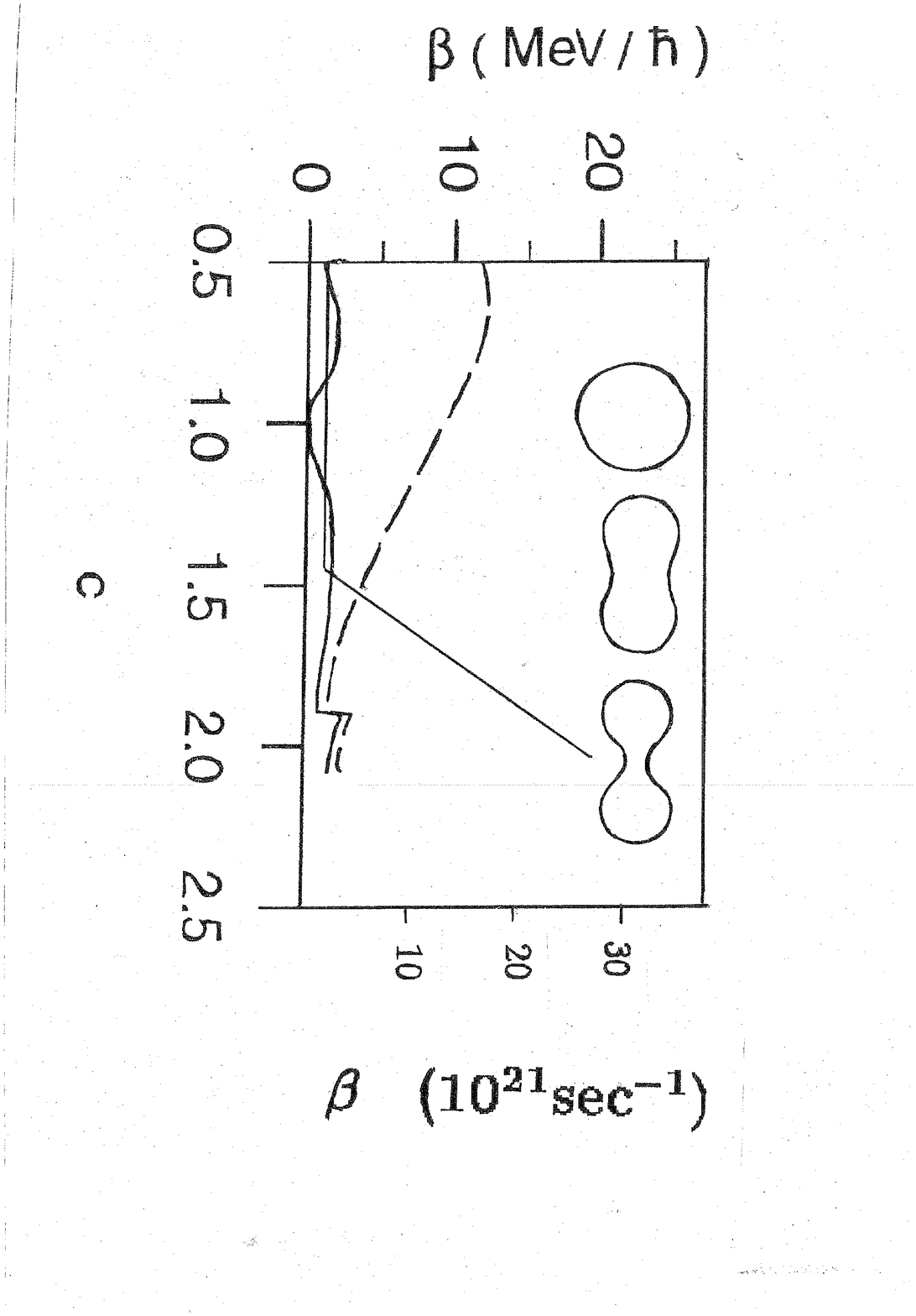}
\protect
\vspace{-0.5cm}
\caption{Comparison of the wall friction (dashed line), the chaos-weighted wall
friction \cite{Pal96}(solid curve),
and the universal phenomenological friction form factor of Fr\"obrich and
Gontchar \cite{FGM93,FG98}(straight lines).}
\end{center}
\end{figure}
function of the elongation coordinate in comparison to the full wall friction.
A comparison is also made with the phenomenological form factor of Fr\"obrich
et al. \cite{FGM93}. Using this form factor
in an overdamped one-dimensional Langevin equation reproduces in a systematic
way the
data on fission (respectively survival) probabilities and pre-scission neutron
multiplicities for light and heavy systems: $^{178}W, ^{188}Pt, ^{200}Pb,
^{213}Fr, ^{224}Th$ and $^{251}Es$.
For compact shapes, both form factors are of the same order of magnitude. For
light systems, it is essential to have this relatively weak friction in order
to reproduce the data. This has been confirmed by Pal et al. \cite{Pal96} by
performing Langevin calculations with their input for the systems above.
However,  they cannot reproduce the neutron multiplicities for $^{251}Es$,
because their form factor does not increase for large deformations. In Ref.
\cite{FGM93} this rise was introduced in order to reproduce the pre-scission
neutron multiplicities for very heavy systems with a long saddle to scission
path, whereas the rise turns out to be insensitive to the neutron
multiplicities
of light systems with a short saddle to scission path. The strength at scission
($\beta=30*10^{21}sec^{-1}$)  is comparable to the value of the surface
friction
model in the exit channel of a deep-inelastic collision which looks similar to
the final stage of fission.

\section*{4. On the temperature dependence of the friction form factor}

A reduction of the wall friction is also obtained by Aleshin \cite{A03}, who
derived a modification of the wall formula by relaxing the randomization
assumption and taking into account interactions of the single-particle motion
in dressing \cite{A99} the single-particle propagator by introducing the
spreading width as in linear response theory. In this way, a temperature
dependence of the friction form factor is introduced. This form factor is
calculated for $^{208}Pb$ in the temperature range of $T=2.1-3.3 MeV$ and
compared in Fig.3, again with the wall friction and the phenomenological form
factor of Fr\"obrich et al. \cite{FGM93,FG98}. The temperature dependent form
factor
behaves similarly to the phenomenological one. It is weak at compact shapes
and rises
with increasing deformation but not as strong as the phenomenological one.
\begin{figure}[htb]
\begin{center}
\protect
\includegraphics*[bb= 50 190 460 780,
angle=-270,clip=true,width=11cm]{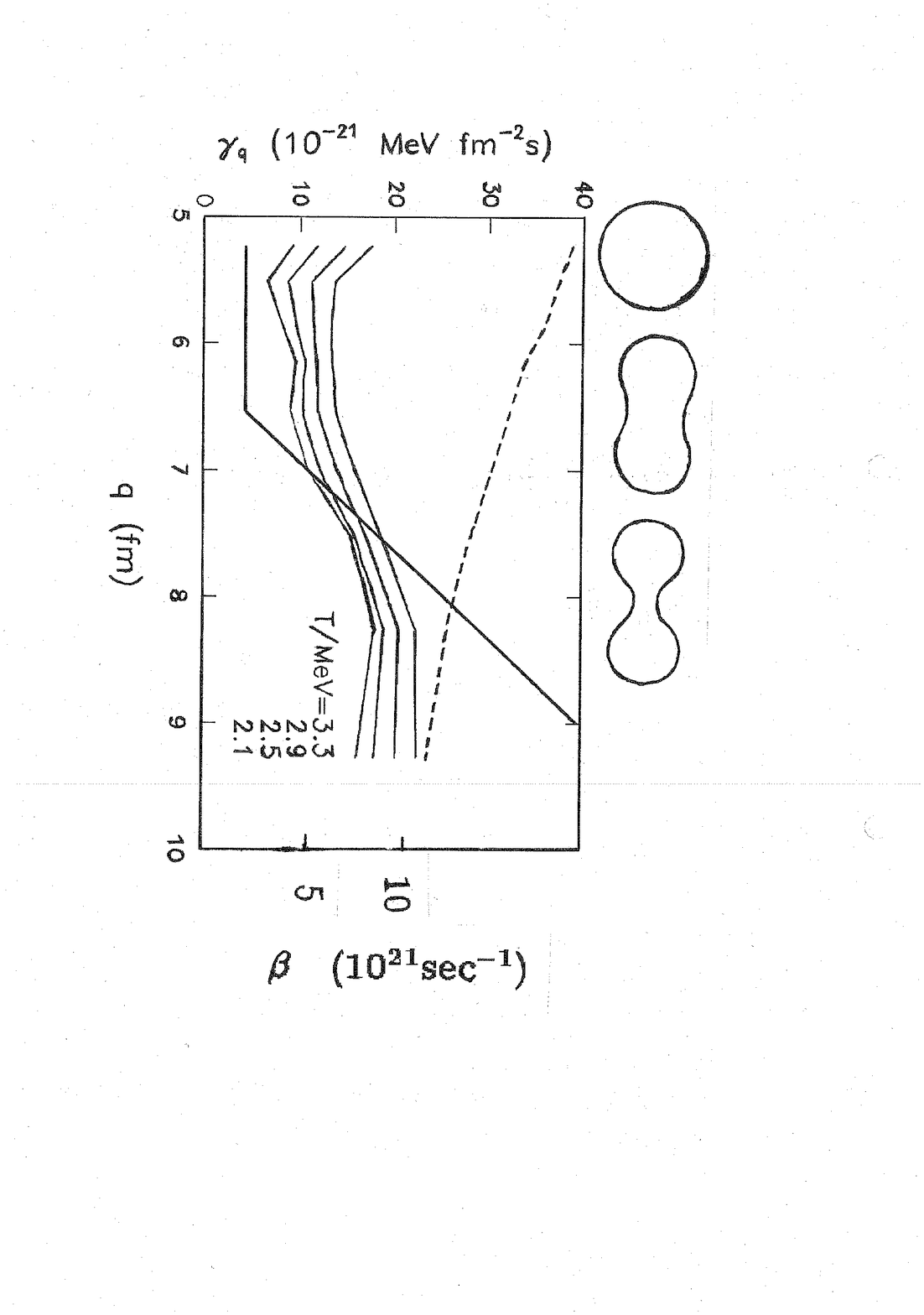}
\protect
\vspace{-0.5cm}
\caption{Comparison of the wall friction (dashed line), the temperature
dependent friction \cite{A03} (solid curves),
and the phenomenological friction form factor of Fr\"obrich and Gontchar
\cite{FGM93,FG98}(straight lines).}
\end{center}
\end{figure}

For a comparable temperature range, the form factor of Aleshin looks
 similar to
that of Ivanyuk et al.\cite{IHPY97}, which is also weaker than the wall
friction for compact shapes and also does not show a steep rise for large
elongation. It would be desirable to perform Langevin calculations coupled to
an evaporation procedure using the temperature- and deformation-dependent form
factors of Refs.\cite{A03,IHPY97}. A first step in this direction was done in
Ref.\cite{KIP02}.

The necessity for clarifying the role of the deformation and
temperature dependence is exemplified in a recent paper by Dioszegi et al.
\cite{DSM02}
who were able to reproduce their data with a modified statistical model
(containing a number of uncertain parameters)
by applying either a temperature dependent friction form factor (with a
stronger temperature dependence than that discussed above) or with a
deformation-dependent form factor which is in accordance with the
phenomenological form
factor weak at compact shapes and stronger at large deformations.

A side remark: The temperature dependence of the friction form factor seems
to be clearer in the decay of metallic clusters, where the viscosity of the
bulk materials is measured. There, one is in the hydrodynamical regime
(two-body
viscosity), and friction becomes weaker with increasing
temperature \cite{F01}.
\section*{5. Remarks on quantal corrections}
In principle, one should use  quantum
mechanically calculated transport coefficients instead of
phenomenological input.
Quantum mechanics is not only hidden in the phenomenological parameters, but
also influences e.g. the fission rates beyond the classical Langevin results
up to quite high temperatures. A fission rate calculated with an influence
functional path integral technique gives e.g. a 20\% enhancement as compared to
a Kramers rate for fission of $^{224}Th$ at a temperature of $1.57
MeV$ \cite{FT92}. At lower temperatures, e.g. when dealing with Langevin models
for superheavy element formation, quantum effects are even more important. For
instance, when calculating rates around the so called cross over temperature,
one has to apply more complicated techniques, e.g. those of Ref.\cite{RA02}.

\section*{6. Conclusions}

The main purpose of the present contribution was to discuss properties of the
friction form factor necessary to describe data concerning fission of hot
nuclei. Arguments for a
modification of the wall formula were collected. To be consistent with a large
variety of data
friction needs to be comparably weak at compact configurations and stronger
for elongated shapes. The role of the temperature dependence of the friction
form factor predicted by microscopic theory has still to be clarified by using
it in Langevin calculations and confronting the results with experimental data.
In the present paper, we discussed only systems where fission follows the
formation of an equilibrated compound system. However, there is an increasing
amount of data in which contributions of fast fission are identified; i.e.
there is a need for modelling the fast fission process. A systematic analysis
of
as many data as possible should be done with the same model, if possible within
a multi-dimensional Langevin dynamics to which an evaporation procedure for
light-particle emission is coupled.
In this way it may be possible to decide upon the friction formfactor with
respect to its deformation (and temperature) dependence, and finally  to arrive
at a unified picture for fission of hot nuclei.


\end{document}